\pdfoutput=1
\documentclass[acmsmall,screen]{acmart}

\setcopyright{none}
\copyrightyear{2024}
\acmYear{2024}
\acmDOI{}

\usepackage{fancyhdr}
\usepackage[capitalize]{cleveref}
\usepackage{siunitx}
\usepackage[frozencache]{minted}  

\setminted{fontsize=\small}
\newmintinline[haskell]{haskell}{}
\newmintinline[racket]{racket}{}

\newcommand{\miniKanren}{\textsc{miniKanren}}
\newcommand{\typedKanren}{\textsc{typedKanren}}
\newcommand{\muKanren}{\textsc{$\upmu$Kanren}}
\newcommand{\fasterMinikanren}{\href{https://github.com/michaelballantyne/faster-minikanren}{\texttt{faster-minikanren}}}
\newcommand{\OCanren}{\href{https://ocanren.readthedocs.io}{\textsc{OCanren}}}
\newcommand{\klogic}{\href{https://github.com/UnitTestBot/klogic}{\textsc{klogic}}}
\newcommand{\canrun}{\href{https://github.com/tgecho/canrun_rs?tab=readme-ov-file}{\texttt{canrun\_rs}}}

\newcommand{\logict}{\href{https://hackage.haskell.org/package/logict}{\texttt{logict}}}
\newcommand{\unificationfd}{\href{https://hackage.haskell.org/package/unification-fd}{\texttt{unification-fd}}}
\newcommand{\Molog}{\href{https://github.com/acfoltzer/Molog}{\texttt{Molog}}}

\newcommand{\matche}{\textbf{match}$^e$}
\newcommand{\conde}{\textbf{cond}$^e$}

\newcommand{\totalLib}{\href{https://hackage.haskell.org/package/total}{\texttt{total}}}
\newcommand{\lensLib}{\href{https://hackage.haskell.org/package/lens}{\texttt{lens}}}
\newcommand{\criterionLib}{\href{https://hackage.haskell.org/package/criterion}{\texttt{criterion}}}

\AtBeginDocument{%
  }

\begin{document}

\title{\typedKanren{}: Statically Typed Relational Programming with Exhaustive Matching in Haskell}

\author{Nikolai Kudasov}
\email{n.kudasov@innopolis.ru}
\orcid{0000-0001-6572-7292}
\affiliation{%
  \institution{Innopolis University}
  \city{Innopolis}
  \state{Tatarstan Republic}
  \country{Russia}
}

\author{Artem Starikov}
\email{a.starikov@innopolis.university}
\orcid{0009-0000-9896-7655}
\affiliation{%
  \institution{Innopolis University}
  \city{Innopolis}
  \state{Tatarstan Republic}
  \country{Russia}
}

\renewcommand{\shortauthors}{Kudasov and Starikov}

\settopmatter{printacmref=false}
\settopmatter{printfolios=true}
\renewcommand\footnotetextcopyrightpermission[1]{}
\pagestyle{fancy}
\fancyfoot{}
\fancyfoot[R]{miniKanren'24}
\fancypagestyle{firstfancy}{
  \fancyhead{}
  \fancyhead[R]{miniKanren'24}
  \fancyfoot{}
}
\makeatletter
\let\@authorsaddresses\@empty
\makeatother

\begin{abstract}
  We present 
  a statically typed
  embedding of relational programming (specifically a dialect of \miniKanren{} with disequality constraints)
  in Haskell. Apart from handling types, our dialect extends standard relational
  combinator repertoire with a variation of relational matching that supports static exhaustiveness checks.
  To hide the boilerplate definitions and support comfortable logic programming
  with user-defined data types we use generic programming via \texttt{GHC.Generics}
  as well as metaprogramming via Template Haskell.
  We demonstrate our dialect on several examples and compare its performance
  against some other known implementations of \miniKanren{}.
\end{abstract}




\maketitle
\thispagestyle{firstfancy}

\section{Introduction}
\label{sec:introduction}

Haskell is a statically typed non-strict purely functional programming language that
offers immense flexibility when it comes to support of programming paradigms.
Indeed, while maintaining a functional core, its \haskell{do}-notation,
which is overloaded via the \haskell{Monad} type class, provides an excellent
environment for different kinds of imperative\footnote{Here, by ``imperative'' we mean programs that consist of sequences of instructions.
These may, but need not support mutable state, exceptions, nondeterminism, and so on.}
programming. Additionally, Haskell offers a variety of generic programming tools~\cite{HinzeJones2001,SheardJones2002}
which are often used to automate the boilerplate for the user when implementing domain-specific languages.
The \haskell{do}-notation (as well as monad comprehensions~\cite{Wadler1990})
can also be overloaded to offer logic programming capabilities (such as in \logict~\cite{KiselyovShanFriedmanSabry2005}).
However, unlike logic programming languages, Haskell does not provide
unification variables or any support for unification of values out of the box.

Relational programming~\cite{TheReasonedSchemer2005} is a kind of logic programming
that relies on building programs as multiway relations. This approach allows
using such relations as functions in different directions, depending on
which of the arguments are known. A notable example is program synthesis
from a relational interpreter~\cite{ByrdHolkFriedman2012}.

\miniKanren{}, introduced\footnote{Although core ideas of \miniKanren{} can be traced to earlier work~\cite{TheReasonedSchemer2005,KiselyovShanFriedmanSabry2005}.} by Byrd in his PhD thesis~\cite{Byrd2009},
is a family\footnote{See a list of \miniKanren{} implementations at \url{http://minikanren.org/\#implementations}.} of relational programming languages embedded into other languages.
Most \miniKanren{} implementations are untyped or unityped, meaning that all
terms have the same type, regardless of what the terms represent. This follows
the uniform representation approach popular in LISP-like languages as well as Prolog.

Almost all\footnote{The only exception we know of is \Molog{} by Adam~C.~Foltzer which is an unfinished project with a similar approach to the one presented in this paper.}
\miniKanren{} embeddings in Haskell that we know of also do not
make use of the expressive type system that Haskell has to offer.
Thus, we are aiming to fill in the gap and offer a
complete statically typed \miniKanren{} embedding in Haskell.

\subsection{Related Work}

Standalone typed logic programming languages such as Mercury and Curry exist.
Remarkably, both languages exhibit an ML-style type system, very reminiscent of Haskell.
A notable feature of Mercury's type system is the \emph{uniqueness types}
that are used, in particular, to enable effects such as input/output.
\typedKanren{}, being embedded in Haskell, inherits many useful type system features.
Although we do not yet explicitly allow effects in \typedKanren{},
we believe our \haskell{Goal} monad can be easily extended into a monad transformer,
similarly to \haskell{LogicT}~\cite{KiselyovShanFriedmanSabry2005}.

There are typed embeddings of \miniKanren{} into OCaml (\OCanren{}~\cite{KosarevBoulytchev2018}),
Kotlin~(\klogic{}~\cite{klogic2023}), and Rust (\canrun{}~\cite{canrun-rs}).
To the best of our knowledge, none of these implementations provide a typed version of \matche{}.
\typedKanren{} provides a typed version of regular matching as well a version with \emph{static} exhaustiveness checks.




\subsection{Contribution}

In this paper, we present an embedding of \miniKanren{} into Haskell.
Our dialect, named \typedKanren{}, enables typed relational programming
in Haskell with typed unification. Specifically, our contribution is as follows:

\begin{enumerate}
  \item In~\cref{sec:core}, we design the relational engine at the core of \typedKanren{}.
  For the underlying representation we follow the classical \miniKanren{} implementations like \muKanren~\cite{HemannFriedman2013},
  however, we also set up a system of type classes that enables typed relational programming on the user side.
  \item In~\cref{sec:exhaustive-matching}, we implement matching, discuss differences
  between functional and relational matching. Here we also introduce a new combinator for
  \emph{exhaustive} relational matching, enabling better safety and performance for some
  relational programs.
  \item In~\cref{sec:metaprogramming}, we use generic programming and metaprogramming techniques
  to hide most of the boilerplate away from the user, so that user-defined data types
  could be easily used in a relational program.
  \item In~\cref{sec:performance}, we compare performance of our implementation
  against other \miniKanren{} implementations,
  including \fasterMinikanren{}~\cite{faster-miniKanren},
  \OCanren{}~\cite{KosarevBoulytchev2018}, and \klogic{}~\cite{klogic2023}.
\end{enumerate}

The implementation of \typedKanren{} is available on GitHub at
\href{https://github.com/SnejUgal/typedKanren}{github.com/SnejUgal/typedKanren}.
A separate repository with combined benchmarks for \fasterMinikanren{}, \OCanren{}, \klogic{}, and \typedKanren{}
is available at \href{https://github.com/SnejUgal/typedKanren-benchmarks}{github.com/SnejUgal/typedKanren-benchmarks}.

\section{Core}
\label{sec:core}

In this section, we introduce the core of \typedKanren{} and describe some design and implementation choices.

From the user perspective, \typedKanren{} consists of
\begin{enumerate}
  \item The \haskell{Logical} typeclass that specifies the types
  that may enter the relational world; specifically, these
  are the types that have their logical counterparts,
  support unification, and conversion between data representations for functional and relational programs.
  \item The \haskell{Goal} monad, providing the context for
  the relational programs, together with the goal combinators,
  including unification, disunification (disequality), conditional and matching operators,
  and fresh variable generators.
  \item Automation utilities that help making most user-defined types \haskell{Logical}.
\end{enumerate}

Ignoring the types, our implementation follows closely the classical implementations
of \miniKanren{}~\cite{ByrdHolkFriedman2012,HemannFriedman2013}. That said,
our implementation is in Haskell, a non-strict (or ``lazy'') functional language,
which means that we do not have to rely on the explicitly constructed thunks,
making the core implementation somewhat more direct, in our opinion.


\subsection{Logical Types and Unification Terms}

The objects of relational programming in \typedKanren{} are \emph{logical} types.
Values of such types may admit unification variables in places where a regular Haskell type
would always have a regular specified value. For example, consider the following recursive parametrically polymorphic type of trees:

\begin{minted}{haskell}
data Tree a
  = Empty                   -- ^ An empty tree.
  | Leaf a                  -- ^ A leaf holding a value.
  | Node (Tree a) (Tree a)  -- ^ A node with two subtrees.
\end{minted}

Here, the \haskell{Leaf} constructor has an associated value (a field) of type \haskell{a}
and \haskell{Node} has two fields of type \haskell{Tree a}.
When going into the relational world, we might expect the tree to be underspecified,
having an undetermined unification variable in place of one or both of the branches.
Therefore, relational programming demands a similar, but separate data type, a relational counterpart to \haskell{Tree}:

\begin{minted}{haskell}
data LogicTree a
  = LogicEmpty
  | LogicLeaf (Term a)
  | LogicNode (Term (Tree a)) (Term (Tree a))
\end{minted}

The type constructor \haskell{Term} provides the logical counterpart to its argument,
except the whole term is also allowed to be a unification variable:

\begin{minted}{haskell}
data Term a
  = Var !(VarId a)   -- ^ A unification variable.
  | Value !(Logic a) -- ^ A logical version of type a.
\end{minted}

The variable identifiers \haskell{VarId} are parametrized with a phantom type parameter
to avoid accidental confusion between unification variables for terms of different types.
Under the hood, a variable identifier is merely a machine-sized integer:

\begin{minted}{haskell}
newtype VarId a = VarId Int
\end{minted}

The type family \haskell{Logic} maps types to their relational counterparts.
For example, \haskell{Logic (Tree a)} is the same as \haskell{LogicTree a}.
To keep track and allow extending this correspondence, we introduce the \haskell{Logical} typeclass:

\begin{minted}{haskell}
class Logical a where
  type Logic a = r | r -> a
  unify :: Logic a -> Logic a -> State -> Maybe State
  walk :: State -> Logic a -> Logic a
  occursCheck :: VarId b -> Logic a -> State -> Bool
  inject :: a -> Logic a
  extract :: Logic a -> Maybe a
\end{minted}

Although the typeclass formally requires a type family and five methods,
our main focus lies only with the type family, since the methods normally
have standard implementation which is derived automatically via \haskell{GHC.Generics}~\cite{HinzeJones2001},
an automation process described in more detail in~\cref{sec:generic-logical}.

Technically, the logical counterparts to user-defined types may also be generated
automatically with Template Haskell~\cite{SheardJones2002}, as is described in~\cref{sec:th-logical}.
That said, while the methods of \haskell{Logical} are normally used in the background,
the logical types face the user, in particular when dealing with type errors.

Note that in the definition of type family \haskell{Logic} we impose a functional dependency,
which ensures that the original type \haskell{a} can always be uniquely recovered
by the compiler if it knows the logical type. This is required, since most of the methods
deal only with the \haskell{Logic} types and without the functional dependency
the instance of \haskell{Logical} may become ambiguous.

Now, as we have established above, \haskell{Tree a} has its logical counterpart,
so we can implement\footnote{We omit the implementation of methods here, see \cref{sec:generic-logical} for the details of generic implementations for them.}
an instance of \haskell{Logical}:

\begin{minted}{haskell}
instance Logical a => Logical (Tree a) where
  type Logic (Tree a) = LogicTree a
  ...
\end{minted}

It is important that the type parameter \haskell{a} is also \haskell{Logical},
since the definition of \haskell{LogicTree} makes use of \haskell{Term a},
which in turn relies on \haskell{Logic a}, which is only defined when \haskell{a} is \haskell{Logical}.

\subsection{Unification of Terms}

The main job of the relational engine of \typedKanren{} is to keep track of
what is known about the terms in a relational program (possibly in different parallel branches).
This information is tracked in the state of the engine and consists of
the substitution for unification variables and the disequality constraints.

Following classical \miniKanren{} implementations~\cite{ByrdHolkFriedman2012,HemannFriedman2013},
we implement ``triangular'' substitutions~\cite[\S 2.2.6]{BaaderSnyderHandbook00} with an ``occurs'' check
in the unification procedure. We rely on integers to represent unification variables,
and use efficient purely functional integer maps~\cite{OkasakiGill1998} to represent substitutions.

In our implementation, we keep unification variables for all types in a single map.
To achieve that, we use existential types to temporarily ``forget'' the type of a term\footnote{In this definition,
we preserve the knowledge that \haskell[fontsize=\footnotesize]{a} is \haskell[fontsize=\footnotesize]{Logical}.
While this is not critical for the unification, it plays a role in the implementation of disequality constraints.}:

\begin{minted}{haskell}
data ErasedTerm where
  ErasedTerm :: Logical a => Term a -> ErasedTerm
\end{minted}

After forgetting the types, we can keep all substitutions in one place:

\begin{minted}{haskell}
newtype Subst = Subst (IntMap ErasedTerm)
\end{minted}

Of course, when looking up a variable in a substitution, we no longer
know the type of the term. However, since the variable identifiers facing the user
are always annotated with the proper type (and the user cannot safely coerce),
we are able to hide the only\footnote{A similar lookup is needed in the implementation of disequality constraints,
but the two can be refactored to only require a single call to \haskell[fontsize=\footnotesize]{unsafeCoerce}, retaining all other properties.}
unsafe coercion and wrap it into a safe function:

\begin{minted}{haskell}
lookupSubst :: VarId a -> Subst -> Maybe (Term a)
lookupSubst = unsafeCoerce IntMap.lookup
\end{minted}

The remaining implementation of the unification is a straightforward adaptation
of the standard approach~\cite[\S D.3]{ByrdHolkFriedman2012}, so we omit it here, referring
the reader to the source code of modules
\href{https://github.com/SnejUgal/typedKanren/blob/master/src/Kanren/Core.hs}{\texttt{Kanren.Core}} and
\href{https://github.com/SnejUgal/typedKanren/blob/master/src/Kanren/Goal.hs}{\texttt{Kanren.Goal}}.

The disequality constraints follow \miniKanren{}~\cite[\S D.1]{ByrdHolkFriedman2012}.
Each constraint is bound to a single unification variable, and the overall set of constraints
is represented with the following data type:

\begin{minted}{haskell}
newtype Disequalities = Disequalities (IntMap [(ErasedTerm, Subst)])
\end{minted}

Again, we provide a safe extraction that uses unsafe coercion under the hood.
Otherwise, the implementation of disequality constraints is standard.

\subsection{The Goal Monad}

Relational programs in \typedKanren{} consist largely of programming with \emph{goals},
essentially in the same way as in other \miniKanren{} implementations.
However, since Haskell provides great syntactical support for imperative-looking code
via the \haskell{do}-notation, it is very convenient to upgrade goals to a proper monadic context.

We define the \haskell{Goal x} to be the function that when given an initial state
generates a stream of possible states (satisfying the relations of the goal)
each with a value of type \haskell{x} associated with it:

\begin{minted}{haskell}
newtype Goal x = Goal {runGoal :: State -> Stream (State, x)}
\end{minted}

We define the ``immature streams''~\cite[\S 4.2]{HemannFriedman2013} here in a standard way,
except relying on an algebraic data type with lazy constructors instead of explicit thunks:

\begin{minted}{haskell}
data Stream a
  = Done
  | Yield a (Stream a)  -- "mature" stream
  | Await (Stream a)    -- "immature" stream
\end{minted}

Naturally, \haskell{Stream} and \haskell{Goal} possess the structure of a monad.
We implement the corresponding \haskell{Monad} instances, following the standard definition~\cite[\S D.1]{ByrdHolkFriedman2012},
relying on the interleaving of streams.

The \haskell{State} simply contains information about substitutions,
disequality constraints, and a global counter used to generate fresh unification variables:

\begin{minted}{haskell}
data State = State
  { knownSubst    :: !Subst
  , disequalities :: !Disequalities
  , maxVarId      :: !Int
  }
\end{minted}

We provide the following basic goal constructors:
\begin{enumerate}
  \item \haskell{successo :: x -> Goal x} is a goal that always succeeds.
  This is the same as \haskell{return} (from the \haskell{Monad} typeclass).
  \item \haskell{failo :: Goal x} is a goal that always fails.
  \item \haskell{(===) :: Logical a => Term a -> Term a -> Goal ()} is a goal
  that unifies the two given terms of type \haskell{a}.
  \item \haskell{conj :: Goal x -> Goal y -> Goal y} represents the conjunction
  of two goals. The result of the first goal is ignored\footnote{Hence, the type of the first goal can be different from the second goal and the result type.},
  while the result of the second goal is kept. This corresponds
  to \haskell{(>>)} (from the \haskell{Monad} typeclass), and more general versions
  of conjunction include \haskell{ap} and \haskell{>>=}. A version of \haskell{conj}
  that conjuncts a list of goals is \haskell{conjMany}.
  \item \haskell{disj :: Goal x -> Goal x -> Goal x} represents the disjunction
  of two goals. The streams from the two goals are interleaved. A version of \haskell{disj}
  that disjuncts a list of goals is \haskell{disjMany}.
  \item \haskell{conde :: [[Goal ()]] -> Goal ()} is \typedKanren{}'s version of \conde{}:
  the outer list represents alternatives (disjunctions), and inner lists represent conjunctions.
  \item \haskell{(=/=) :: Logical a => Term a -> Term a -> Goal ()} is a goal
  that disunifies the two given terms of type \haskell{a}.
  \item \haskell{fresh :: Fresh v => Goal v} is a goal that produces a fresh unification variable
  of a given type.
\end{enumerate}

Regarding \haskell{fresh}, the user is not expected to implement any new instances of \haskell{Fresh} typeclass.
This is because there exists an instance for logical terms as well as tuples of logical terms:

\begin{minted}{haskell}
instance (Logical a) => Fresh (Term a) where ...
instance (Logical a, Logical b) => Fresh (Term a, Term b) where ...
\end{minted}

Finally, to run the relational program, we execute the goal:

\begin{minted}{haskell}
run :: Fresh v => (v -> Goal ()) -> [v]
\end{minted}

This function produces a \emph{lazy} list of values (usually of type \haskell{Term a} for some \haskell{a}).

\subsection{Example}

With all the basic ingredients in place, we can implement a simple relational program.
First, consider a functional program that extracts all leaf values from a \haskell{Tree}:

\begin{minted}{haskell}
leaves :: Tree a -> [a]
leaves t = case t of
  Empty -> []
  Leaf x -> [x]
  Node l r -> leaves l ++ leaves r
\end{minted}

Converting this into a relational program we get

\begin{minted}{haskell}
leaveso :: Logical a => Term (Tree a) -> Term [a] -> Goal ()
leaveso t xs = disjMany
  [ do
      t === Value LogicEmpty
      xs === Value LogicNil
  , do
      x <- fresh
      t === Value (LogicLeaf x)
      xs === Value (LogicCons x (Value LogicNil))
  , do
      (l, r, as, bs) <- fresh
      t === Value (LogicNode l r)
      leaveso l as
      leaveso r bs
      appendo as bs xs
  ]
\end{minted}

In this example, we use \haskell{disjMany} with a list of \haskell{do}-blocks,
as we see it less syntactically noisy compared to \haskell{conde} in \typedKanren{}.
The constructors \haskell{LogicNil} and \haskell{LogicCons} are the
logical counterparts to the empty and non-empty list constructors\footnote{It is possible
to rely on \haskell[fontsize=\footnotesize]{OverloadedLists} extension to use regular syntax for lists,
but we prefer to use explicit constructors here for clarity of presentation.}.

It is possible to run this relational program in both directions.
First, we may specify the tree and ask \typedKanren{} to compute all possible lists
corresponding to the leaves of a given tree (we expect exactly one possibly):
\begin{minted}{haskell}
>>> t = Node (Node (Leaf 1) Empty) (Leaf (2 :: Int))
>>> run (leaveso (inject' t))
[Value (LogicCons (Value 1) (Value (LogicCons (Value 2) (Value LogicNil))))]
\end{minted}

We may extract the non-logical value(s) (safely, using \haskell{Maybe}):
\begin{minted}{haskell}
>>> extract' <$> run (leaveso (inject' t))
[Just [1,2]]
\end{minted}

Specifying the list and leaving the tree as an argument, we effectively
run the relation ``backwards''. Here, we take and print the first five\footnote{Since
some leaves may be \haskell[fontsize=\footnotesize]{Empty}, there are infinitely many trees
corresponding to the list \haskell[fontsize=\footnotesize]{[1, 2]}.}
trees that have two leaves with values \haskell{1} and \haskell{2} (in that order):

\begin{minted}{haskell}
>>> mapM_ (print . extract') $ take 5 (run (`leaveso` (inject' [1, 2 :: Int])))
Just (Node (Leaf 1) (Leaf 2))
Just (Node Empty (Node (Leaf 1) (Leaf 2)))
Just (Node (Leaf 1) (Node Empty (Leaf 2)))
Just (Node (Leaf 1) (Node (Leaf 2) Empty))
Just (Node (Node Empty (Leaf 1)) (Leaf 2))
\end{minted}

\section{Exhaustive Matching}
\label{sec:exhaustive-matching}

In his thesis~\cite{Byrd2009}, Byrd introduces a logical matching operator \matche{},
a relational counterpart to the functional matching operators
\textbf{pmatch}~\cite[Appendix~B]{Byrd2009} and \textbf{dmatch}~\cite[Appendix~C]{ByrdHolkFriedman2012}.
Of course, \typedKanren{} also provides \haskell{matche}, however, in this section,
we explore the design space for statically typed relational matching operators,
considering both safety and efficiency.

\subsection{Prisms as First-Class Patterns in Haskell}

Unlike LISP, quasiquotation is not built into Haskell\footnote{Quasiquotation is supported by Template Haskell,
however, it is unclear how to properly use it for patterns here,
since we want the main combinators to be regular Haskell functions,
not Template Haskell functions (which would correspond to LISP macros).
So, we leave research into feasibility of quasiquotation for future work.},
and thus patterns and data constructors only work in one direction at a time:
when used in a \haskell{case}-expression or in function argument the data constructor
can be used to match against a Haskell value, while when used in an expression, the same
data constructor is used to create a Haskell value.

In a relational setting, we want to work with patterns that work simultaneously in both
directions, relying on unification under the hood. To achieve this, we use the well-known
technique for first-class patterns in Haskell~--- the \emph{prisms}.
Semantically, a simple prism of type \haskell{Prism' s a} is a Haskell value that
is equivalent to a pair of functions for matching (of type \haskell{s -> Maybe a}) and constructing values (of type \haskell{a -> s}).
The two common representations for prisms are the van~Laarhoven representation~\cite{vanLaarhoven2009},
used in the Kmett's \lensLib{} library~\cite{lens}, and the profunctor optics representation~\cite{PickeringGibbonsWu2017}.
The former is widely used in Haskell, so we go with it in this paper.
However, we do not see any reason that profunctor optics would not work just as well in this setting.

An important generalization for prisms (as well as other optics),
is a separation of types for matching and construction. Specifically, a prism of type \haskell{Prism s t a b}
semantically consists of a matching function of type \haskell{s -> Either t a} and a construction function of type \haskell{b -> t}.
This separation of types is important as it allows changing types when modifying a value under a prism
or enforce further properties on the prisms.

Below we present two examples, relevant to the design of typed relational matching in \typedKanren{}.
Both examples will rely on the following definition and relevant prisms\footnote{We omit the implementation of prisms, as only the types are important in the discussion and implementation may vary depending on the chosen representation.}:

\begin{minted}{haskell}
data Result a b = Ok a | Fail b

_Ok   :: Prism (Result a c) (Result b c) a b
_Fail :: Prism (Result c a) (Result c b) a b
\end{minted}

\subsubsection{Changing Types with Prisms}

In the following example, we use the prism \haskell{_Fail}
to modify the type of errors from \haskell{Err} to \haskell{String}
by using a standard Haskell function \haskell{show} to convert a value into a string:

\begin{minted}{haskell}
example :: Result Int Err -> Result Int String
example result = result
  & _Ok   %~ (+1) -- increment result if it is Ok
  & _Fail %~ show -- convert error into a String if it is Fail
\end{minted}

\subsubsection{Eliminating Alternatives with Prisms}

In this next example, we use the ability of the prism \haskell{Ok}
to change the type of its corresponding constructor to any other type
when it does not appear in the value we are matching against:

\begin{minted}{haskell}
example :: Result Int Bool -> Int
example result =
  case matching _Ok result :: Either Void Bool of
    -- successful matching on Ok, extracting n :: Int
    Right n -> n
    -- matching failed, x :: Either Void Bool
    Left x -> case x of
      -- Fail is still a possible alternative
      Fail b -> fromEnum b
      -- Ok is impossible here, hence the use of absurd
      Ok n   -> absurd n
\end{minted}

\subsection{Typed Relational Matching}

The prism-based relational matching is presented in \typedKanren{} with two combinators.

First, \haskell{matche :: Term a -> Goal ()} is the default matching combinator without any pattern matching branches.

Second, \haskell{on} combinator acts as a modifier, adding a single pattern matching branch
to an existing matching; the type of \haskell{on} is fairly straightforward:

\begin{minted}{haskell}
on :: (Logical a, Fresh v)
  => Prism' (Logic a) v -- ^ The pattern presented as a prism.
  -> (v -> Goal x)      -- ^ The handler for this one alternative.
  -> (Term a -> Goal x) -- ^ Matching for other alternatives.
  -> (Term a -> Goal x) -- ^ Extended matching (+1 alternative).
\end{minted}

Technically, all branches here have the same type and could be organized in a list,
resembling more the traditional \matche{}. However, we choose the approach of
modifiers for matchers, as it generalizes well for the safer exhaustive matching
variant discussed below, following the design of Gonzalez's \totalLib{} library~\cite{Gonzalez2015,total}.

As an example, consider a relational version of list concatenation in \typedKanren{}:

\begin{minted}{haskell}
-- the following prisms are used to match on lists in a relational program
_LogicNil  :: Prism' (Logic [a]) ()
_LogicCons :: Prism' (Logic [a]) (a, Term [a])

appendo :: Logical a => Term [a] -> Term [a] -> Term [a] -> Goal ()
appendo xs ys zs = xs & (matche         -- matching on xs
  & on _LogicNil (\() -> ys === zs)     -- if xs is empty, then unify ys and zs
  & on _LogicCons (\(x, xs') -> do      -- if xs is non-empty with head x and tail xs'
      zs' <- fresh                      -- then
      zs === Value (LogicCons x zs')    -- zs is also nonempty with head x and tail zs'
      appendo xs' ys zs'))              -- such that xs' ++ ys unifies with zs'
\end{minted}

It might be useful to compare it with a functional version\footnote{We use \haskell[fontsize=\footnotesize]{let}-bindings for some (sub)expressions to underline the similar computational structure to the relational version.}
(working in one direction only):

\begin{minted}{haskell}
append :: [a] -> [a] -> [a]
append xs ys = case xs of
  [] -> let zs = ys in zs
  (x:xs') ->
    let zs = x : zs'
        zs' = append xs' ys
    in zs
\end{minted}

We note that the structure of the program is similar:
\begin{itemize}
  \item \haskell{matche} and \haskell{on} replace the \haskell{case}-expression in the relational version;
  \item \haskell{fresh} and binding in a \haskell{do}-block is used instead of \haskell{let}
  \item in the relational program, the order of goals in the \haskell{_LogicCons}-branch
  matters while the order of \haskell{let}-bindings in the functional program does not;
  while there are approaches to fair relational conjunction~\cite{KiselyovShanFriedmanSabry2005,LuMaFriedman2019,LozovBoulytchev2020},
  \typedKanren{} does not implement those variants.
\end{itemize}

Although we have achieved typed relational matching, which works well for many occasions,
we believe, it presents a few disadvantages that can be eliminated, at least for some relational programs:
\begin{enumerate}
  \item Under the hood, each matching branch contributes a separate runtime-matching,
  which is akin to a series of nested \haskell{if}-expressions. This is, of course,
  suboptimal: our \haskell{matche} has (under the hood) a \haskell{case}-expression
  per matching branch, whereas the functional version only has one overall \haskell{case}-expression
  for the entire match. To the best of our knowledge, other implementations of \miniKanren{},
  including the statically typed \OCanren{}, rely on unification and \conde{}, similarly resulting
  in multiple redundant matches on the input value.
  \item Many relational programs, similarly to \haskell{appendo} above, still consider
  the full set of distinct (non-overlapping) patterns in the matching branches.
  At the same time, in Haskell, and many other statically typed languages, exhaustiveness checks for
  the pattern matching (e.g. in \haskell{case}-expressions) are considered a vital resource
  for making sure no important branch is forgotten, especially when refactoring the code.
  Therefore, it appears to be practically useful to allow users enable exhaustiveness checking
  for relational matching.
\end{enumerate}

\subsection{Exhaustive Relational Matching}

In this section, we describe a variation of relational matching with
exhaustiveness checks, enabling safer relational programs.

Since we use prisms as first-class patterns for the relational matching,
we cannot rely on the built-in exhaustiveness checker in Haskell,
since it operates only with Haskell's native patterns, not prisms.
Gonzalez~\cite{Gonzalez2015,total} has proposed a mechanism for exhaustive
matching that works with lenses, prisms, and traversals. In \typedKanren{},
we are following along the same ideas, except it becomes slightly more involved
since we are using prisms in both directions.

The main idea is to introduce a version of \haskell{on} that visibly reduces
the possible alternatives for matching tracked in the types. A first approximation is as follows:

\begin{minted}{haskell}
on1 :: (Logical a, Fresh v)
  => Prism (Logic a) (Logic b) v Void
  -> (v -> Goal x)
  -> (Term b -> Goal x)
  -> (Term a -> Goal x)
\end{minted}

The generalized form of the prism ensures that the type \haskell{Logic b} is exactly
a version of the type \haskell{Logic a} where an alternative containing \haskell{v}
is impossible (replaced with \haskell{Void}).
A specialized version of this is exactly what is used in Gonzalez's \totalLib~library~\cite{total}.

Unfortunately, using \haskell{Void} in the last type parameter of the prism
effectively forbids using this prism for construction. Indeed, the construction
function extracted from such a prism would have the type \haskell{Void -> Logic b},
which is impossible to apply while staying in the safe Haskell territory.

To preserve the ability to construct values and keep track of the remaining alternatives,
in \typedKanren{} we use the type-level annotations via the \haskell{Tagged} wrapper type:

\begin{minted}{haskell}
-- | A value of type b, tagged with type a.
newtype Tagged a b = Tagged b
\end{minted}

\subsubsection{Patterns Tagged with Exhaustiveness Information}

In the context of exhaustive relational matching, the type-level tags carry
the information about checked and remaining cases.
To keep track of this information, we use special version of the logical prisms
for the tagged logical values.


For example, such annotated prism for
the \haskell{Ok} constructor of \haskell{Result} has the following type and is defined as a coercion over the regular prism,
imposing zero runtime cost\footnote{Since \haskell[fontsize=\footnotesize]{Tagged} is a \haskell[fontsize=\footnotesize]{newtype},
its runtime representation is identical to the underlying type and
corresponding coercion functions are identities that we trust
are optimized away by the Glasgow Haskell Compiler.
For guaranteed zero cost abstraction, \haskell[fontsize=\footnotesize]{unsafeCoerce} can be (safely) used here.}:

\begin{minted}{haskell}
_LogicOk' :: Prism
  (Tagged (ok, fail) (Logic (Result a b)))
  (Tagged (ok', fail) (Logic (Result a b)))
  (Tagged ok (Term a))
  (Tagged ok' (Term a))
_LogicOk' = from _Tagged . _LogicOk . _Tagged
\end{minted}

Let us dissect the type arguments of this prism:
\begin{enumerate}
  \item \haskell{Tagged (ok, fail) (Logic (Result a b))} corresponds to the
  matched logical value of type \haskell{Result a b}. The tags \haskell{ok} and \haskell{fail}
  designate the information about checked (sub)cases for the data constructors \haskell{Ok} and \haskell{Fail}
  respectively. For example, when \haskell{(ok, fail)} is \haskell{(Remaining, Checked)},
  that would mean that only the \haskell{Ok} case remains to be checked.

  \item \haskell{Tagged (ok', fail) (Logic (Result a' b))} corresponds to the
  logical value for the remaining cases (\emph{after} considering the \haskell{Ok} case).
  The tag \haskell{ok'} corresponds to the updated information about the \haskell{Ok} case.
  If matching concerns the entire \haskell{Ok} constructor (regardless of the subcases for the type \haskell{a}),
  then we would expect \haskell{ok'} to be exactly \haskell{Checked}. However, the
  type signature of \haskell{_LogicOk'} allows \haskell{ok'} to be partially checked,
  in case of nested prisms.

  \item \haskell{Tagged ok (Term a)} corresponds to the type of the logical term\footnote{Note that it also includes the case for just a unification variable, which is indicated by the use for \haskell[fontsize=\footnotesize]{Term} instead of \haskell[fontsize=\footnotesize]{Logic}.}
  for the chosen case \haskell{before} handling this alternative.

  \item \haskell{Tagged ok' (Term a')} corresponds to the type of the logical term
  for the chosen case \haskell{after} handling this alternative.
  Note that both the tag and the type are allowed to be changed (since \haskell{Result} is parametrically polymorphic in the contents of its \haskell{Ok} constructor).
\end{enumerate}

For conciseness, we introduce the \haskell{ExhaustivePrism} type alias. The type
of \haskell{_LogicOk'} then becomes:

\begin{minted}{haskell}
_LogicOk' :: ExhaustivePrism
  (Logic (Result a b)) (ok, fail) (ok, fail')
  (Term a) ok ok'
\end{minted}

\subsubsection{Exhaustive Relational Matching}

We define three combinators for exhaustive matching.

First, we define \haskell{on'} which is similar to \haskell{on},
but makes use of annotated prisms. Its type is more complicated:
\begin{minted}{haskell}
on'
  :: (Logical a, Fresh v)
  => ExhaustivePrism (Logic a) m m' v Remaining Checked
  -- ^ The pattern presented as an annotated prism.
  -> (v -> Goal x) -- ^ The handler for this one alternative.
  -> (Matched m' a -> Goal x) -- ^ Matching for other alternatives.
  -> (Matched m a -> Goal x)  -- ^ Extended matching.
\end{minted}

\haskell{on'} specializes the prism so that its case is marked as
\haskell{Checked}. The changed list of tags is then passed to other
alternatives. The \haskell{Remaining} tag demands this case not be
checked previously; although we could allow checking the same pattern
twice, this would pose inconvenient edge cases in nested matching.
\haskell{Matched m a} is an alias for \haskell{Tagged m (Term a)}.

Second, we define \haskell{matche'} which is similar
to \haskell{matche}, but also performs exhaustiveness check by
constraining \haskell{m} to contain only \haskell{Checked} tags (by imposing \haskell{Exhausted m} constraint):
\begin{minted}{haskell}
matche' :: Exhausted m => Matched m a -> Goal x
\end{minted}

Finally, we define \haskell{enter'} which is a
new combinator attaching tags to the term being matched.
\begin{minted}{haskell}
enter' :: (Matched m a -> Goal x) -> Term a -> Goal x
\end{minted}

As an example, consider the following relational program. For clarity, we show
tags (\haskell{m}) passed to each combinator.

\begin{minted}{haskell}
resulto :: (Logical a, Logical b) => Term (Result a b) -> Goal ()
resulto r = r & (matche'              -- (  Checked,   Checked)
  & on' _LogicOk' (\_ -> successo ())   -- (Remaining,   Checked)
  & on' _LogicFail' (\_ -> successo ()) -- (Remaining, Remaining)
  & enter')
\end{minted}

Now consider what happens when an alternative is omitted. If so, its
corresponding tag is never instantiated to a particular type and remains a type
variable.

\begin{minted}{haskell}
resulto :: (Logical a, Logical b) => Term (Result a b) -> Goal ()
resulto r = r & (matche'                -- (ok,   Checked)
  & on' _LogicFail' (\_ -> successo ()) -- (ok, Remaining)
  & enter')
\end{minted}

The constraint \haskell{Exhausted m} imposed by \haskell{matche'} will reduce
to \haskell{Exhausted ok}. However, since \haskell{ok} is a type variable not
constrained by anything, this constraint will not be satisfied and the program
will fail to compile.


\section{Hiding the Boilerplate}
\label{sec:metaprogramming}

Here we explain how the boilerplate definitions, helper functions, and class instances
can be generated with \texttt{GHC.Generics}~\cite{HinzeJones2001} and Template Haskell~\cite{SheardJones2002}.

\subsection{Generating the Logical Types}
\label{sec:th-logical}

For each Haskell data type, there may exist many logical variants,
depending on where we allow unification variables to occur. By default,
and in the examples in the previous sections, we have allowed
unification variables to occur anywhere. More specifically, for every data constructor,
each of the fields of type \haskell{T} in the original data type becomes
a field of type \haskell{Term T} in the relational version of that type.
This is a very straightforward approach, that maximizes the use of unification variables,
and can be easily automated.

We use Template Haskell~\cite{SheardJones2002} to generate such \emph{maximal} logical types
together with the appropriate instances. For mutually recursive types, we also provide
more customizable functions to separately generate types and instances.
Specifically, we provide the following template functions:
\begin{minted}{haskell}
makeLogic      :: Name -> Q [Dec]   -- type and instances
makeLogicType  :: Name -> Q [Dec]   -- only type
makeLogicTypes :: [Name] -> Q [Dec] -- mutually recursive types

-- generate a default Generic-based Logical instance(s)
makeLogicalInstance :: Name -> Name -> Q [Dec]
makeLogicalInstances :: [(Name, Name)] -> Q [Dec]
\end{minted}

Given the name of a user-defined data type, we generate its maximal logical counterpart,
as well as a generic \haskell{Logical} instance. For example, consider the following user
code defining a \haskell{Tree} data type and invoking \haskell{makeLogic} for it:

\begin{minted}{haskell}
{-# LANGUAGE TemplateHaskell #-}
...
data Tree a
  = Leaf a
  | Node (Tree a) (Tree a)
  deriving (Eq, Show, Generic)
makeLogic ''Tree
\end{minted}

Our Template Haskell code systematically inspects the user-defined data type declaration
and generates the following code:

\begin{minted}{haskell}
data LogicTree a
  = LogicLeaf (Term a)
  | LogicNode (Term (Tree a)) (Term (Tree a))
  deriving (Generic)

instance Logical a => Logical (Tree a) where
  type Logic (Tree a) = LogicTree a
  unify = genericUnify
  subst = genericSubst
  occursCheck = genericOccursCheck
  inject = genericInject
  extract = genericExtract
\end{minted}

The \haskell{generic*} functions are explained in detail in \cref{sec:generic-logical}.
Importantly, \haskell{makeLogic} supports the vast majority of valid Haskell data type definitions,
including records, infix constructors, strictness annotations\footnote{Since both constructors
of \haskell[fontsize=\footnotesize]{Term a} are defined as strict, preserving strictness
annotations in logical versions of user-defined constructors does not alter the intended strictness
of terms.},
and generalized algebraic data types. It should be noted though, that generalized algebraic data types
might not have \haskell{GHC.Generic} instances. Support for TH-generated implementations
of \haskell{Logical} methods for such cases is not present and is subject to future work.

For our code to work, the following language extensions are required on the user side in addition
to \haskell{TemplateHaskell}: \haskell{DeriveGeneric}, \haskell{TypeFamilies}.

Implementation of \haskell{} makes the following implicit decisions that are not configurable at the moment, pending future work:
\begin{enumerate}
  \item the type name and all (prefix) constructor names are prefixed with \haskell{Logic}: \haskell{Foo} becomes \haskell{LogicFoo};
  \item field names are prefixed with \haskell{logic} and capitalized: \haskell{fooBar} becomes \haskell{logicFooBar};
  \item all fields are wrapped in \haskell{Term}: \haskell{T} becomes \haskell{Term T}.
\end{enumerate}

\subsection{Generic Unification for Logical Types}
\label{sec:generic-logical}

To enable unification for the logical types, \typedKanren{} requires the following methods
of the \haskell{Logical} typeclass to be implemented:
\begin{enumerate}
  \item \haskell{unify} performs unification of two logical values;
  \item \haskell{walk} updates a logical value, applying known substitutions
  to any unification variables in the logical value;
  \item \haskell{occursCheck} inspects a logical value to see if a given
  variable is present in one of its subterms;
  \item \haskell{inject} converts regular value of type \haskell{a} into
  a logical value of type \haskell{Logic a} (without any variables);
  \item \haskell{extract} attempts to convert a logical value of type \haskell{Logic a} to
  the regular value of type \haskell{a}; of course, since it is not possible
  in presence of unification variables, the result is wrapped in a \haskell{Maybe}.
\end{enumerate}

Fortunately, all of these methods follow one of the two standard implementations:
\begin{enumerate}
  \item For base types, such as \haskell{Int} or \haskell{Bool} for which
  their logical counterpart coincides with the regular type, unification
  simplifies to a mere equality check, \haskell{walk} and \haskell{inject} are
  identities, \haskell{occursCheck} is always false (there are no unification variables allowed),
  and \haskell{extract} simply wraps input in a \haskell{Just} constructor.

  \item For cases when logical type is distinct, their structure is expected
  to match: constructor-to-constructor and field-to-field. In this case,
  we can rely on the generic structure of algebraic types (as provided by \haskell{GHC.Generics})
  to derive the corresponding methods. This approach is described below in detail.
\end{enumerate}

We introduce a generic counterpart to \haskell{Logical} typeclass. The idea is
to implement all necessary methods for a generic representation and then,
for user-defined types, implement each method via conversion back and forth between the generic representation
and the user code. Importantly, relying on the generic representation and such conversions
is a zero-cost abstraction in GHC~\cite{HinzeJones2001}, providing a safer (compared to Template Haskell)
and efficient generic implementation for the \haskell{Logical} methods.

The generic logical typeclass takes two parameters: the representation types of
the user-defined type and its logical counterpart.
We found the multiparameter definition to be somewhat easier to work with here,
although it is technically possible to use an associated type family as in \haskell{Logical}.
Note the use of \haskell{Proxy d} to ensure we always know which type \haskell{f}
the method corresponds to. This is the alternative to the functional dependency used
in the definition of the associated type family \haskell{Logic}.

\begin{minted}{haskell}
class GLogical f f' where
  gunify :: Proxy f -> f' p -> f' p -> State -> Maybe State
  gwalk :: Proxy f -> State -> f' p -> f' p
  goccursCheck :: Proxy f -> VarId b -> f' p -> State -> Bool
  ginject :: f p -> f' p
  gextract :: f' p -> Maybe (f p)
\end{minted}

For the generic implementation, it is sufficient to consider the following cases:
\begin{enumerate}
  \item \haskell{V1} — representation for an empty type;
  \item \haskell{U1} — representation for a unit type (think of a constructor without fields);
  \item \haskell{f :*: g} — representation for a product of types (think of a constructor with at least one field);
  \item \haskell{f :+: g} — representation for a sum of types (think of a data type with at least one constructor);
  \item \haskell{K1 i c} — representation for a field of type \haskell{c};
  \item \haskell{M1 i t f} — same as \haskell{f}, except with some metadata over it.
\end{enumerate}

To illustrate the generic implementation, we demonstrate the implementation of \haskell{gunify} for the specified cases above.
The empty and unit types are trivial, as there is nothing to check, and we can say that values of these types
always unify without any changes to the state:
\begin{minted}{haskell}
instance GLogical V1 V1 where gunify _ _ _ = Just
instance GLogical U1 U1 where gunify _ _ _ = Just
\end{minted}

For the sum of types, we check if the same alternative is present in the two input values.
If they are the same, then we recursively descend, otherwise, unification fails:
\begin{minted}{haskell}
instance (GLogical f f', GLogical g g')
    => GLogical (f :+: g) (f' :+: g') where
  gunify _ (L1 x) (L1 y) = gunify (Proxy @f) x y
  gunify _ (R1 x) (R1 y) = gunify (Proxy @g) x y
  gunify _ _ _           = const Nothing
\end{minted}

For products, we start by unifying first components, and then, if successful,
we unify the second components:
\begin{minted}{haskell}
instance (GLogical f f', GLogical g g')
    => GLogical (f :*: g) (f' :*: g') where
  gunify _ (x1 :*: y1) (x2 :*: y2) state = do
    state' <- gunify (Proxy @f) x1 x2
    gunify (Proxy @g) y1 y2 state'
\end{minted}

When encountering a field of type \haskell{c} in the regular type,
we expect \haskell{Term c} in its logical counterpart,
and appeal to unification for \haskell{c}:
\begin{minted}{haskell}
instance (Logical c)
    => GLogical (K1 i c) (K1 i' (Term c)) where
  gunify _ (K1 x) (K1 y) = unify' x y
\end{minted}

Note that none of the instances above explicitly deal with the unification variables.
This is handled by the \haskell{unify'} function that is defined in the core for \haskell{Term} and follows the
standard \miniKanren{} implementation for unification~\cite{ByrdHolkFriedman2012,HemannFriedman2013}.
We provide the implementation here for context:

\begin{minted}{haskell}
unify' :: Logical a => Term a -> Term a -> State -> Maybe State
unify' l r state =
  case (shallowWalk state l, shallowWalk state r) of
    (Var x, Var y)
      | x == y -> Just state
    (Var x, r')
      | occursCheck' x r' state -> Nothing
      | otherwise -> addSubst x r' state
    (l', Var y)
      | occursCheck' y l' state -> Nothing
      | otherwise -> addSubst y l' state
    (Value l', Value r') -> unify l' r' state
\end{minted}

Finally, we provide an instance for \haskell{M1} where we simply skip metadata:
\begin{minted}{haskell}
instance (GLogical f f')
    => GLogical (M1 i t f) (M1 i' t' f') where
  gunify _ (M1 x) (M1 y) = gunify (Proxy @f) x y
\end{minted}

With the generic instances in place, we provide an implementation for any
type that has a generic representation:

\begin{minted}{haskell}
genericUnify
  :: forall a.
    (Generic (Logic a), GLogical (Rep a) (Rep (Logic a)))
  => Logic a
  -> Logic a
  -> State
  -> Maybe State
genericUnify l r = gunify (Proxy @(Rep a)) (from l) (from r)
\end{minted}

The implementation of \haskell{genericWalk}, \haskell{genericOccursCheck}, \haskell{genericInject},
and \haskell{genericExtract} follows the same pattern. Complete implementation is available in the
\href{https://github.com/SnejUgal/typedKanren/blob/master/src/Kanren/GenericLogical.hs}{Kanren.GenericLogical}
module.

While generic implementation works well in most situations, there are
some limitations, since not all Haskell data types provide a generic implementation.
In particular, GHC cannot derive \haskell{Generic} instance for some fairly simple
generalized algebraic data types (GADTs). While there do exist approaches to generic programming
that works better with GADTs~\cite{SerranoMiraldo2018}, they are not part of the standard \haskell{GHC.Generics} toolkit,
so we leave such support for future work.

\section{Performance Evaluation}
\label{sec:performance}

\begin{figure*}
  \begin{center}
  \begin{tabular}{| l || c | c | c | c | c |}
  \hline
    & \haskell{exp}, ms & \haskell{log}, ms & \haskell{quines}, ms & \haskell{twines}, ms & \haskell{thrines}, ms \\ [0.5ex]
    \hline\hline
    \fasterMinikanren{}~\cite{faster-miniKanren} (Racket) & $106.5$ & $17.6$ & $294.8$ & $261.4$ & $487.0$ \\
    \hline
    \OCanren{}~\cite{KosarevBoulytchev2018} & $463.7$ & $61.4$ & $777.3$ & $717.6$ & $1\,258.9$ \\
    \hline
    \klogic{}~\cite{klogic2023} & $942.2$ & $87.8$ & $1\,071.6$ & $1\,260.5$ & $5\,071.0$ \\
    \hline
    \typedKanren{} (this paper) & $588.1$ & $68.1$ & $1\,084.0$ & $1\,580.0$ & $3\,775.0$ \\
    \hline
  \end{tabular}
  \end{center}
  \caption{Preliminary benchmark results.
  The benchmarks were executed on an ASUS ZenBook 15 with an AMD Ryzen 7 7735U CPU, 16GB of memory, using NixOS, Linux 6.10.2.}
  \label{fig:benchmarks}
\end{figure*}

To evaluate the performance of \typedKanren{}, we have implemented

\begin{enumerate}
  \item A relational arithmetic system, based on a binary representation~\cite{KiselyovByrdFriedmanShan2008}.
  The system includes the exponentiation and integer logarithm relations, which are used in the benchmarks.
  The implementation is available in the module \href{https://github.com/SnejUgal/typedKanren/blob/master/src/Kanren/Data/Binary.hs}{Kanren.Data.Binary}.
  \item A relational Scheme interpreter~\cite{ByrdHolkFriedman2012}.
  Running the interpreter ``backwards'' allows systematically searching for programs (S-expressions)
  that produce the desired result. In particular, such interpreter can be used to generate quines — programs that evaluate to themselves.
  Quine generation is used in the benchmarks. The implementation is available in the module \href{https://github.com/SnejUgal/typedKanren/blob/master/src/Kanren/Data/Scheme.hs}{Kanren.Data.Scheme}.
\end{enumerate}

Preliminary benchmarks are shown in~\cref{fig:benchmarks}:
\begin{enumerate}
  \item \haskell{exp} — computing $3^5$ using the relational arithmetic system;
  \item \haskell{log} — computing the integer logarithm $\log_3 243$ using the relational arithmetic system;
  \item \haskell{quines} — generation of 100 quines via the relational Scheme interpreter;
  \item \haskell{twines} — generation of 15 twines\footnote{A \emph{twine} is a pair of programs A and B such that A evaluates to B and B evaluates to A.} via the relational Scheme interpreter;
  \item \haskell{thrines} — generation of 2 thrines\footnote{A \emph{thrine} is a triple of programs A, B, and C such that A evaluates to B, B evaluates to C, and C evaluates to A.} via the relational Scheme interpreter.
\end{enumerate}

To benchmark the Haskell implementation, we use \criterionLib{}~\cite{criterion},
a widely-used Haskell library for performance measurement and analysis of Haskell functions.
To ensure proper computation we force evaluation of results to normal form (via \haskell{NFData} instances).
For benchmarks of other implementations, we reuse the setup from \klogic{} paper~\cite{klogic2023}.
The code and instructions for replicating the benchmarks for \fasterMinikanren{}, \OCanren{}, \klogic{}, and \typedKanren{}
is available at \href{https://github.com/SnejUgal/typedKanren-benchmarks}{github.com/SnejUgal/typedKanren-benchmarks}.

Based on the benchmark results, we can say that \typedKanren{} performs on par with \klogic{},
but at the moment is outperformed by all other implementations. We believe that the following factors contribute to this result:
\begin{enumerate}
  \item A feature that is shared by LISP-based implementations like \fasterMinikanren{}
  is cheap \haskell{inject}. In fact, in LISP all user values are essentially S-expressions
  and converting a value into one available for relational setting is not necessary.
  Similarly, \OCanren{} implements tagless logical values~\cite[\S 6.2]{KosarevBoulytchev2018},
  also allowing for zero-cost injection. This relies on polymorphic unification~\cite[\S 5]{KosarevBoulytchev2018},
  which might be possible in Haskell via another kind of generic programming~\cite{LammelJones2003}.

  \item We do not (yet) implement the \racket{set-var-val!} optimization
  to reduce the cost of looking up variables. This appears to be one of the main
  optimizations in \fasterMinikanren{}~\cite{faster-miniKanren}
  which at the moment is de facto the fastest implementation.
  While mutable variables are normally inaccessible in Haskell,
  we may use the local state~\cite{LaunchburyJones1994} safely inside the \haskell{Goal}
  for unification variables to gain performance.
  A similar optimization is implemented in \unificationfd{} library~\cite{unification-fd}.

  That said, we have conducted some experiments implementing a version of this optimization
  both with \haskell{STRef}\footnote{see \href{https://github.com/snejugal/typedKanren/pull/13}{github.com/snejugal/typedKanren/pull/13}}
  and \haskell{IORef}\footnote{see \href{https://github.com/snejugal/typedKanren/pull/13}{github.com/snejugal/typedKanren/pull/14}},
  none of which affected the performance of \typedKanren{} positively.

  \item At the moment we rarely use strictness annotations,
  and thus our implementation may accumulate too many unwanted thunks.
  We should perform proper strictness analysis to enhance performance and
  avoid excessive memory usage.

  \item To ensure that the terms are fully evaluated, we rely on \haskell{NFData}.
  However, this might have a negative effect on the performance in the cases where
  the term is already fully evaluated. Although we do not believe that this contributes much,
  we should properly analyze the effect of forcing in the future.
\end{enumerate}

\section{Conclusion and Future Work}
\label{sec:conclusion}

We present a working embedding of relational programming
in a typed non-strict functional programming language Haskell.
Our implementation, \typedKanren{}, is a feature full dialect of \miniKanren{} with unification, disequality constraints,
and static typing support.
\typedKanren{} is capable of expressing many classical relational programs,
including relational arithmetic systems~\cite{KiselyovByrdFriedmanShan2008} and a relational Scheme interpreter~\cite{ByrdHolkFriedman2012},
which we implement in full and use in our benchmarks.

To provide relational matching, we rely on prisms, an implementation of first-class patterns in Haskell.
In addition to the typed version of \matche{}, we discuss potential shortcomings of relational matching
and provide a matching operator with exhaustiveness checking.

To assist the user with development of relational programs in presence of user-defined types,
we provide generic programming and metaprogramming tools to generate logical counterparts to Haskell types
and derive necessary instances to provide unification capabilities.
For relational matching, we rely on existing metaprogramming tools provided by the \lensLib~library~\cite{lens}
to generate regular prisms and also provide a Template Haskell function to generate exhaustive prisms for logical types.

Unlike \OCanren{}~\cite{KosarevBoulytchev2018}, we do not provide any quasiquotation support
to allow nicer syntax for relational programs. The main reason is that \typedKanren{} is an active
work-in-progress, and we prefer to exhaust design choices within Haskell syntax before
leaning on quasiquotation with a separate parser. That said, support for quasiquotation
is a prominent direction for future work.

Another useful form of automation is conversion of existing functional programs
into relational ones~\cite{LozovVyatkinBoulytchev2018}. Indeed, using Template Haskell
it appears feasible to perform such transformations automatically for many Haskell functions.

On the performance side, \typedKanren{} at the moment is underperforming (although not critically) compared
with \fasterMinikanren{} (the fastest known untyped implementation), \OCanren{}, and \klogic{}.
There are a number of well-known optimizations that should be properly implemented in the core of \typedKanren{},
achieving performance that is on par or exceeds the competitive implementations.
The first contender for a significant improvement is the Haskell analog of \racket{set-var-val!} optimization.
Other ideas for optimizations and heuristics speeding up substitution may be taken from Wren Romano's \unificationfd{} library.

\begin{acks}
We thank Nikolay Shilov for his comments on an earlier draft of this paper.
We thank the anonymous reviewers of miniKanren'24 workshop for their valuable
feedback on an earlier draft of this paper.
\end{acks}

\bibliographystyle{ACM-Reference-Format}
\bibliography{ms}


\begin{thebibliography}{28}


\ifx \showCODEN    \undefined \def \showCODEN     #1{\unskip}     \fi
\ifx \showDOI      \undefined \def \showDOI       #1{#1}\fi
\ifx \showISBNx    \undefined \def \showISBNx     #1{\unskip}     \fi
\ifx \showISBNxiii \undefined \def \showISBNxiii  #1{\unskip}     \fi
\ifx \showISSN     \undefined \def \showISSN      #1{\unskip}     \fi
\ifx \showLCCN     \undefined \def \showLCCN      #1{\unskip}     \fi
\ifx \shownote     \undefined \def \shownote      #1{#1}          \fi
\ifx \showarticletitle \undefined \def \showarticletitle #1{#1}   \fi
\ifx \showURL      \undefined \def \showURL       {\relax}        \fi
\providecommand\bibfield[2]{#2}
\providecommand\bibinfo[2]{#2}
\providecommand\natexlab[1]{#1}
\providecommand\showeprint[2][]{arXiv:#2}

\bibitem[{Baader} and {Snyder}(2001)]%
        {BaaderSnyderHandbook00}
\bibfield{author}{\bibinfo{person}{F. {Baader}} {and} \bibinfo{person}{W.
  {Snyder}}.} \bibinfo{year}{2001}\natexlab{}.
\newblock \showarticletitle{Unification Theory}.
\newblock In \bibinfo{booktitle}{\emph{Handbook of Automated Reasoning}},
  \bibfield{editor}{\bibinfo{person}{J.A. {Robinson}} {and}
  \bibinfo{person}{A.~{Voronkov}}} (Eds.). Vol.~\bibinfo{volume}{I}.
  \bibinfo{publisher}{Elsevier Science Publishers}, \bibinfo{pages}{447--533}.
\newblock


\bibitem[Ballantyne(2015)]%
        {faster-miniKanren}
\bibfield{author}{\bibinfo{person}{Michael Ballantyne}.}
  \bibinfo{year}{2015}\natexlab{}.
\newblock \bibinfo{booktitle}{\emph{A fast implementation of miniKanren with
  disequality and absento, compatible with Racket and Chez}}.
\newblock
\urldef\tempurl%
\url{https://github.com/michaelballantyne/faster-minikanren}
\showURL{%
\tempurl}


\bibitem[Byrd(2009)]%
        {Byrd2009}
\bibfield{author}{\bibinfo{person}{William~E. Byrd}.}
  \bibinfo{year}{2009}\natexlab{}.
\newblock \emph{\bibinfo{title}{Relational Programming in miniKanren:
  Techniques, Applications, and Implementations}}.
\newblock \bibinfo{thesistype}{Ph.\,D. Dissertation}. \bibinfo{address}{USA}.
\newblock Advisor(s) Friedman, Daniel P.
\newblock
\showISBNx{9781109504682}
\newblock
\shownote{AAI3380156}.


\bibitem[Byrd et~al\mbox{.}(2012)]%
        {ByrdHolkFriedman2012}
\bibfield{author}{\bibinfo{person}{William~E. Byrd}, \bibinfo{person}{Eric
  Holk}, {and} \bibinfo{person}{Daniel~P. Friedman}.}
  \bibinfo{year}{2012}\natexlab{}.
\newblock \showarticletitle{miniKanren, Live and Untagged: Quine Generation via
  Relational Interpreters (Programming Pearl)}. In
  \bibinfo{booktitle}{\emph{Proceedings of the 2012 Annual Workshop on Scheme
  and Functional Programming}} (Copenhagen, Denmark)
  \emph{(\bibinfo{series}{Scheme '12})}. \bibinfo{publisher}{Association for
  Computing Machinery}, \bibinfo{address}{New York, NY, USA},
  \bibinfo{pages}{8–29}.
\newblock
\showISBNx{9781450318952}
\urldef\tempurl%
\url{https://doi.org/10.1145/2661103.2661105}
\showDOI{\tempurl}


\bibitem[Gonzalez({[n.\,d.]})]%
        {total}
\bibfield{author}{\bibinfo{person}{Gabriella Gonzalez}.}
  \bibinfo{year}{[n.\,d.]}\natexlab{}.
\newblock \bibinfo{booktitle}{\emph{\texttt{total} library}}.
\newblock
\urldef\tempurl%
\url{https://hackage.haskell.org/package/total}
\showURL{%
\tempurl}


\bibitem[Gonzalez(2015)]%
        {Gonzalez2015}
\bibfield{author}{\bibinfo{person}{Gabriella Gonzalez}.}
  \bibinfo{year}{2015}\natexlab{}.
\newblock \bibinfo{booktitle}{\emph{\texttt{total-1.0.0}: Exhaustive pattern
  matching using traversals, prisms, and lenses}}.
\newblock
\urldef\tempurl%
\url{https://www.haskellforall.com/2015/01/total-100-exhaustive-pattern-matching.html}
\showURL{%
\tempurl}


\bibitem[{Hemann} and {P. Friedman}(2013)]%
        {HemannFriedman2013}
\bibfield{author}{\bibinfo{person}{Jason {Hemann}} {and}
  \bibinfo{person}{Daniel {P. Friedman}}.} \bibinfo{year}{2013}\natexlab{}.
\newblock \showarticletitle{$\upmu$Kanren: A minimal functional core for
  relational programming}. In \bibinfo{booktitle}{\emph{Proceedings of the 2013
  Workshop on Scheme and Functional Programming}}.
\newblock
\urldef\tempurl%
\url{http://webyrd.net/scheme-2013/papers/HemannMuKanren2013.pdf}
\showURL{%
\tempurl}


\bibitem[Hinze and {Peyton Jones}(2001)]%
        {HinzeJones2001}
\bibfield{author}{\bibinfo{person}{Ralf Hinze} {and} \bibinfo{person}{Simon
  {Peyton Jones}}.} \bibinfo{year}{2001}\natexlab{}.
\newblock \showarticletitle{Derivable Type Classes}.
\newblock \bibinfo{journal}{\emph{Electronic Notes in Theoretical Computer
  Science}} \bibinfo{volume}{41}, \bibinfo{number}{1} (\bibinfo{year}{2001}),
  \bibinfo{pages}{5--35}.
\newblock
\showISSN{1571-0661}
\urldef\tempurl%
\url{https://doi.org/10.1016/S1571-0661(05)80542-0}
\showDOI{\tempurl}
\newblock
\shownote{2000 ACM SIGPLAN Haskell Workshop (Satellite Event of PLI 2000)}.


\bibitem[{Kamenev} et~al\mbox{.}(2023)]%
        {klogic2023}
\bibfield{author}{\bibinfo{person}{Yury {Kamenev}}, \bibinfo{person}{Dmitrii
  {Kosarev}}, \bibinfo{person}{Dmitry {Ivanov}}, \bibinfo{person}{Denis
  {Fokin}}, {and} \bibinfo{person}{Dmitri {Boulytchev}}.}
  \bibinfo{year}{2023}\natexlab{}.
\newblock \showarticletitle{{klogic: miniKanren in Kotlin}}. In
  \bibinfo{booktitle}{\emph{Proceedings of the Fifth miniKanren Workshop
  (miniKanren '23)}}.
\newblock
\urldef\tempurl%
\url{http://minikanren.org/workshop/2023/minikanren23-final4.pdf}
\showURL{%
\tempurl}


\bibitem[Kiselyov et~al\mbox{.}(2008)]%
        {KiselyovByrdFriedmanShan2008}
\bibfield{author}{\bibinfo{person}{Oleg Kiselyov}, \bibinfo{person}{William~E.
  Byrd}, \bibinfo{person}{Daniel~P. Friedman}, {and}
  \bibinfo{person}{Chung-chieh Shan}.} \bibinfo{year}{2008}\natexlab{}.
\newblock \showarticletitle{Pure, Declarative, and Constructive Arithmetic
  Relations (Declarative Pearl)}. In \bibinfo{booktitle}{\emph{Functional and
  Logic Programming}}, \bibfield{editor}{\bibinfo{person}{Jacques Garrigue}
  {and} \bibinfo{person}{Manuel~V. Hermenegildo}} (Eds.).
  \bibinfo{publisher}{Springer Berlin Heidelberg}, \bibinfo{address}{Berlin,
  Heidelberg}, \bibinfo{pages}{64--80}.
\newblock
\showISBNx{978-3-540-78969-7}


\bibitem[Kiselyov et~al\mbox{.}(2005)]%
        {KiselyovShanFriedmanSabry2005}
\bibfield{author}{\bibinfo{person}{Oleg Kiselyov}, \bibinfo{person}{Chung-chieh
  Shan}, \bibinfo{person}{Daniel~P. Friedman}, {and} \bibinfo{person}{Amr
  Sabry}.} \bibinfo{year}{2005}\natexlab{}.
\newblock \showarticletitle{Backtracking, Interleaving, and Terminating Monad
  Transformers (Functional Pearl)}. In \bibinfo{booktitle}{\emph{Proceedings of
  the Tenth ACM SIGPLAN International Conference on Functional Programming}}
  (Tallinn, Estonia) \emph{(\bibinfo{series}{ICFP '05})}.
  \bibinfo{publisher}{Association for Computing Machinery},
  \bibinfo{address}{New York, NY, USA}, \bibinfo{pages}{192–203}.
\newblock
\showISBNx{1595930647}
\urldef\tempurl%
\url{https://doi.org/10.1145/1086365.1086390}
\showDOI{\tempurl}


\bibitem[Kmett({[n.\,d.]})]%
        {lens}
\bibfield{author}{\bibinfo{person}{Edward Kmett}.}
  \bibinfo{year}{[n.\,d.]}\natexlab{}.
\newblock \bibinfo{booktitle}{\emph{\texttt{lens} library}}.
\newblock
\urldef\tempurl%
\url{https://hackage.haskell.org/package/lens}
\showURL{%
\tempurl}


\bibitem[Kosarev and Boulytchev(2018)]%
        {KosarevBoulytchev2018}
\bibfield{author}{\bibinfo{person}{Dmitrii Kosarev} {and}
  \bibinfo{person}{Dmitry Boulytchev}.} \bibinfo{year}{2018}\natexlab{}.
\newblock \showarticletitle{Typed embedding of a relational language in OCaml}.
\newblock \bibinfo{journal}{\emph{arXiv preprint arXiv:1805.11006}}
  (\bibinfo{year}{2018}).
\newblock


\bibitem[Launchbury and Peyton~Jones(1994)]%
        {LaunchburyJones1994}
\bibfield{author}{\bibinfo{person}{John Launchbury} {and}
  \bibinfo{person}{Simon~L. Peyton~Jones}.} \bibinfo{year}{1994}\natexlab{}.
\newblock \showarticletitle{Lazy functional state threads}. In
  \bibinfo{booktitle}{\emph{Proceedings of the ACM SIGPLAN 1994 Conference on
  Programming Language Design and Implementation}} (Orlando, Florida, USA)
  \emph{(\bibinfo{series}{PLDI '94})}. \bibinfo{publisher}{Association for
  Computing Machinery}, \bibinfo{address}{New York, NY, USA},
  \bibinfo{pages}{24–35}.
\newblock
\showISBNx{089791662X}
\urldef\tempurl%
\url{https://doi.org/10.1145/178243.178246}
\showDOI{\tempurl}


\bibitem[Lozov and Boulytchev(2020)]%
        {LozovBoulytchev2020}
\bibfield{author}{\bibinfo{person}{Petr Lozov} {and} \bibinfo{person}{Dmitry
  Boulytchev}.} \bibinfo{year}{2020}\natexlab{}.
\newblock \showarticletitle{On Fair Relational Conjunction}.
\newblock \bibinfo{journal}{\emph{Proceedings of the 2020 miniKanren and
  Relational Programming Workshop}} (\bibinfo{year}{2020}),
  \bibinfo{pages}{1--12}.
\newblock
\urldef\tempurl%
\url{http://minikanren.org/workshop/2020/minikanren-2020-paper1.pdf}
\showURL{%
\tempurl}


\bibitem[Lozov et~al\mbox{.}(2018)]%
        {LozovVyatkinBoulytchev2018}
\bibfield{author}{\bibinfo{person}{Petr Lozov}, \bibinfo{person}{Andrei
  Vyatkin}, {and} \bibinfo{person}{Dmitry Boulytchev}.}
  \bibinfo{year}{2018}\natexlab{}.
\newblock \showarticletitle{Typed Relational Conversion}. In
  \bibinfo{booktitle}{\emph{Trends in Functional Programming}},
  \bibfield{editor}{\bibinfo{person}{Meng Wang} {and} \bibinfo{person}{Scott
  Owens}} (Eds.). \bibinfo{publisher}{Springer International Publishing},
  \bibinfo{address}{Cham}, \bibinfo{pages}{39--58}.
\newblock
\showISBNx{978-3-319-89719-6}


\bibitem[Lu et~al\mbox{.}(2019)]%
        {LuMaFriedman2019}
\bibfield{author}{\bibinfo{person}{Kuang-Chen Lu}, \bibinfo{person}{Weixi Ma},
  {and} \bibinfo{person}{Daniel~P Friedman}.} \bibinfo{year}{2019}\natexlab{}.
\newblock \showarticletitle{Towards a miniKanren with fair search strategies}.
  In \bibinfo{booktitle}{\emph{Proceedings of the 2019 miniKanren and
  Relational Programming Workshop}}. \bibinfo{pages}{1--15}.
\newblock
\urldef\tempurl%
\url{http://minikanren.org/workshop/2019/minikanren19-final1.pdf}
\showURL{%
\tempurl}


\bibitem[Lämmel and Peyton~Jones(2003)]%
        {LammelJones2003}
\bibfield{author}{\bibinfo{person}{Ralf Lämmel} {and} \bibinfo{person}{Simon
  Peyton~Jones}.} \bibinfo{year}{2003}\natexlab{}.
\newblock \showarticletitle{Scrap your boilerplate: a practical approach to
  generic programming}. In \bibinfo{booktitle}{\emph{ACM SIGPLAN International
  Workshop on Types in Language Design and Implementation (TLDI'03)}
  (\bibinfo{edition}{acm sigplan international workshop on types in language
  design and implementation (tldi'03)} ed.)}. \bibinfo{publisher}{ACM Press},
  \bibinfo{pages}{26--37}.
\newblock
\urldef\tempurl%
\url{https://www.microsoft.com/en-us/research/publication/scrap-your-boilerplate-a-practical-approach-to-generic-programming/}
\showURL{%
\tempurl}


\bibitem[Okasaki and Gill(1998)]%
        {OkasakiGill1998}
\bibfield{author}{\bibinfo{person}{Chris Okasaki} {and} \bibinfo{person}{Andrew
  Gill}.} \bibinfo{year}{1998}\natexlab{}.
\newblock \showarticletitle{Fast Mergeable Integer Maps}. In
  \bibinfo{booktitle}{\emph{Workshop on ML}}. \bibinfo{pages}{77--86}.
\newblock
\urldef\tempurl%
\url{https://git.sr.ht/~wklew/containers/blob/b4074eaabf2c2c0f87b0a096a7b4eb3a2f9dee97/papers/Okasaki%20and%20Gill%20-%201998%20-%20Fast%20Mergeable%20Integer%20Maps.pdf}
\showURL{%
\tempurl}


\bibitem[O'Sullivan({[n.\,d.]})]%
        {criterion}
\bibfield{author}{\bibinfo{person}{Bryan O'Sullivan}.}
  \bibinfo{year}{[n.\,d.]}\natexlab{}.
\newblock \bibinfo{booktitle}{\emph{\texttt{criterion}: Robust, reliable
  performance measurement and analysis}}.
\newblock
\urldef\tempurl%
\url{https://hackage.haskell.org/package/criterion}
\showURL{%
\tempurl}


\bibitem[{P. Friedman} et~al\mbox{.}(2005)]%
        {TheReasonedSchemer2005}
\bibfield{author}{\bibinfo{person}{Daniel {P. Friedman}},
  \bibinfo{person}{William {E. Byrd}}, {and} \bibinfo{person}{Oleg
  {Kiselyov}}.} \bibinfo{year}{2005}\natexlab{}.
\newblock \bibinfo{booktitle}{\emph{The Reasoned Schemer}}.
\newblock \bibinfo{publisher}{MIT Press}.
\newblock
\showISBNx{9780262256179}
\urldef\tempurl%
\url{https://doi.org/10.7551/mitpress/5801.001.0001}
\showDOI{\tempurl}


\bibitem[Pickering et~al\mbox{.}(2017)]%
        {PickeringGibbonsWu2017}
\bibfield{author}{\bibinfo{person}{Matthew Pickering}, \bibinfo{person}{Jeremy
  Gibbons}, {and} \bibinfo{person}{Nicolas Wu}.}
  \bibinfo{year}{2017}\natexlab{}.
\newblock \showarticletitle{Profunctor Optics: Modular Data Accessors}.
\newblock \bibinfo{journal}{\emph{Art Sci. Eng. Program.}} \bibinfo{volume}{1},
  \bibinfo{number}{2} (\bibinfo{year}{2017}), \bibinfo{pages}{7}.
\newblock
\urldef\tempurl%
\url{https://doi.org/10.22152/PROGRAMMING-JOURNAL.ORG/2017/1/7}
\showDOI{\tempurl}


\bibitem[Romano({[n.\,d.]})]%
        {unification-fd}
\bibfield{author}{\bibinfo{person}{Wren Romano}.}
  \bibinfo{year}{[n.\,d.]}\natexlab{}.
\newblock \bibinfo{booktitle}{\emph{\texttt{unification-fd}: Simple generic
  unification algorithms}}.
\newblock
\urldef\tempurl%
\url{https://hackage.haskell.org/package/unification-fd}
\showURL{%
\tempurl}


\bibitem[Serrano and Miraldo(2018)]%
        {SerranoMiraldo2018}
\bibfield{author}{\bibinfo{person}{Alejandro Serrano} {and}
  \bibinfo{person}{Victor~Cacciari Miraldo}.} \bibinfo{year}{2018}\natexlab{}.
\newblock \showarticletitle{Generic programming of all kinds}. In
  \bibinfo{booktitle}{\emph{Proceedings of the 11th ACM SIGPLAN International
  Symposium on Haskell}} (St. Louis, MO, USA) \emph{(\bibinfo{series}{Haskell
  2018})}. \bibinfo{publisher}{Association for Computing Machinery},
  \bibinfo{address}{New York, NY, USA}, \bibinfo{pages}{41–54}.
\newblock
\showISBNx{9781450358354}
\urldef\tempurl%
\url{https://doi.org/10.1145/3242744.3242745}
\showDOI{\tempurl}


\bibitem[Sheard and Jones(2002)]%
        {SheardJones2002}
\bibfield{author}{\bibinfo{person}{Tim Sheard} {and}
  \bibinfo{person}{Simon~Peyton Jones}.} \bibinfo{year}{2002}\natexlab{}.
\newblock \showarticletitle{Template meta-programming for Haskell}. In
  \bibinfo{booktitle}{\emph{Proceedings of the 2002 ACM SIGPLAN Workshop on
  Haskell}} (Pittsburgh, Pennsylvania) \emph{(\bibinfo{series}{Haskell '02})}.
  \bibinfo{publisher}{Association for Computing Machinery},
  \bibinfo{address}{New York, NY, USA}, \bibinfo{pages}{1–16}.
\newblock
\showISBNx{1581136056}
\urldef\tempurl%
\url{https://doi.org/10.1145/581690.581691}
\showDOI{\tempurl}


\bibitem[Simmler({[n.\,d.]})]%
        {canrun-rs}
\bibfield{author}{\bibinfo{person}{Erik Simmler}.}
  \bibinfo{year}{[n.\,d.]}\natexlab{}.
\newblock \bibinfo{booktitle}{\emph{\texttt{canrun\_rs}: a Rust logic
  programming library inspired by the *Kanren family of language DSLs.}}
\newblock
\urldef\tempurl%
\url{https://github.com/tgecho/canrun_rs}
\showURL{%
\tempurl}


\bibitem[van Laarhoven(2009)]%
        {vanLaarhoven2009}
\bibfield{author}{\bibinfo{person}{Twan van Laarhoven}.}
  \bibinfo{year}{2009}\natexlab{}.
\newblock \bibinfo{booktitle}{\emph{CPS based functional references}}.
\newblock
\urldef\tempurl%
\url{https://www.twanvl.nl/blog/haskell/cps-functional-references}
\showURL{%
\tempurl}


\bibitem[Wadler(1990)]%
        {Wadler1990}
\bibfield{author}{\bibinfo{person}{Philip Wadler}.}
  \bibinfo{year}{1990}\natexlab{}.
\newblock \showarticletitle{Comprehending monads}. In
  \bibinfo{booktitle}{\emph{Proceedings of the 1990 ACM Conference on LISP and
  Functional Programming}} (Nice, France) \emph{(\bibinfo{series}{LFP '90})}.
  \bibinfo{publisher}{Association for Computing Machinery},
  \bibinfo{address}{New York, NY, USA}, \bibinfo{pages}{61–78}.
\newblock
\showISBNx{089791368X}
\urldef\tempurl%
\url{https://doi.org/10.1145/91556.91592}
\showDOI{\tempurl}


\end{thebibliography}


\end{document}